\documentclass{jaa}
\usepackage{threeparttable}
\usepackage{amsmath}
\usepackage[utf8]{inputenc}
\usepackage{longtable}
\usepackage{subcaption}
\usepackage{lipsum}
\usepackage{graphicx}
\usepackage{natbib}
\usepackage{multirow}
\usepackage[Table,xcdraw]{xcolor}
\usepackage{rotating}

\usepackage{hyperref}
\hypersetup{backref=true,
    pagebackref=true,
    hyperindex=true,
    colorlinks=true,
    breaklinks=true,
    urlcolor= black,
    linkcolor= blue,
    bookmarks=true,
    bookmarksopen=false,
    filecolor=black,
    citecolor=blue,
    linkbordercolor=blue
}

\begin{document}\sloppy

\title{Plans for building a prototype SKA  regional centre in India}
\author{Yogesh Wadadekar\textsuperscript{1}, Dipankar Bhattacharya\textsuperscript{2,5}, Abhirup Datta\textsuperscript{3}, Surajit  Paul\textsuperscript{4} and Divya Oberoi\textsuperscript{1}} 
\affilOne{\textsuperscript{1} National Centre for Radio Astrophysics, TIFR, Post Bag 3, Ganeshkhind, Pune 411007, India\\
\affilTwo{\textsuperscript{2} Inter-University Centre for Astronomy and Astrophysics, Pune 411007, India\\
\affilThree{\textsuperscript{3} Department of Astronomy, Astrophysics and Space Engineering, Indian Institute of Technology Indore, Indore 433552, India.\\}}
\affilFour{\textsuperscript{4} Department of Physics, Savitribai Phule Pune University, Pune 411007, India.\\}}
\affilFive{\textsuperscript{5} Ashoka University, Rajiv Gandhi Education City, Sonipat 131029, India \\}

\twocolumn[{

\maketitle

\corres{yogesh@ncra.tifr.res.in}

\msinfo{---}{---}

\begin{abstract}

In order to deliver the full science potential of the Square Kilometer
Array (SKA) telescope, several SKA Regional Centres (SRCs) will be
required to be constructed in different SKA member countries around
the world. These SRCs will provide high performance compute and
storage for the generation of advanced science data products from the
basic data streams generated by the SKA Science Data Handling and
Processing system, critically necessary to the success of the key
science projects to be carried out by the SKA user community. They
will also provide support to astronomers to enable them to carry out
analysis on very large SKA datasets. Construction of such large data
centres is a technical challenge for all SKA member nations. In such a
situation, each country plans to construct a smaller SRC over the next
few years (2022 onwards), known as a proto-SRC.  In India, we propose
to construct a proto-SRC which will be used for the analysis of data
from SKA pathfinders and precursors with strong Indian involvement
such as uGMRT, Meerkat and MWA. We describe our thinking on some
aspects of the the storage, compute and network of the proto-SRC and
how it will be used for data analysis as well as for carrying out
various simulations related to SKA key science projects led by Indian
astronomers.  We also present our thoughts on how the proto-SRC plans
to evaluate emerging hardware and software technologies and to also
begin software development in areas of relevance to SKA data
processing and analysis such as algorithm implementation, pipeline
development and data visualisation software.

\end{abstract}


}]

\doinum{}
\artcitid{\#\#\#\#}
\volnum{000}
\year{0000}
\pgrange{1--}
\setcounter{page}{1}
\lp{10}


\section{Introduction} 
\label{sec:Introduction}

The SKA Observatory, when fully operational will generate prodigious
quantities of data \citep{Scaife}.  These data will be processed using
certain standardised pipelines within the observatory as a part of the
science data handling and processing. However, this standardised data
processing may not be sufficient for realising the science goals of
different research projects. Some projects may require additional
processing before the data can be used for science, while others may
require some of the processing done at the observatory, to be redone,
with a different software algorithm and/or with a different processing
configuration. Doing all of these things is not at all trivial,
because the volume of data to be ingested, stored and processed will
be in excess of 700 Petabytes per year. The processing capacity
required is about 22 Petaflops. Dedicated network speeds of 100 Gbit/s
will be needed to transport data from the telescopes to the SRCs and
between the SRCs.
 
The SKA will therefore require the integration of
centralised High Performance Computing (HPC) and data analytics
systems into the core of data analysis and visualisation. This
dramatic shift away from the astronomer's desktop (or a small
local cluster) to remotely accessible, large-scale facilities will, in
turn, completely alter the whole ecosystem of astronomy
research. This implies that our traditional data processing methods
will no longer work and the sociology of radio astronomy research will
be irreversibly changed.
 
The resources needed to fully process, distribute, curate and utilise
data flowing from the SKA are currently beyond the scope of the SKA1
construction and operations budget. Recognizing the criticality of
these additional resources for delivering SKA science, the SKAO Data
Flow Advisory Panel recommended in March 2016 that the SKAO Board
encourage SKA member states to form “a collaborative network of SKA
Regional Centres (SRCs) to provide the essential functions that are
not presently provided within the scope of the SKA1 project”.  This
recommendation was soon endorsed by the SKA Board, and thereafter the
SKA organisation formed the SRC Coordination Group (SRCCG) in
mid-2016, with representatives from most SKA member states, including
India, along with external advisors from the Vera C. Rubin Observatory
and CERN. The SRCCG was tasked with defining the principles, policies,
requirements and MoUs for a collaborative network of SRCs. In the
context of the SKA, the term "regional" has been left deliberately
ambiguous. It may refer to a group of countries like those in Western
Europe or to individual large countries like India. What is more
important, is that the regional centres be well distributed across the
globe and designed to cater to local (time zone) users specifically,
while still providing some shared resources for the use of the global
community.

In discussions within the SRCCG, it became quickly obvious, that
although the construction of a large distributed network of data
processing centres was new to astronomy, there were some precedents in
other science domains. Specifically, the Large Hadron Collider project
at CERN, has set up the Worldwide LHC Computing Grid (WLCG), which is
a global collaboration of around 170 computing centres in more than 40
countries, bringing together national and international grid
infrastructures. This tiered grid of data processing centres are
designed to store, distribute and analyse the $\sim50-70$ Petabytes of
data expected every year of operations from the LHC facility. From a
study of the WLCG experience carried out by the SRCCG, it became clear that
the SKAO and the international SKA science community will need to work
collaboratively to shape and establish a shared, distributed data,
computing and networking capability that draws on international
cooperation and enables the full range of SKA science. Such a
distributed and shared capability needs to be persistent (long-lived),
despite depending on funding from a variety of sources from multiple
governments and stakeholders. It also needs to be coherent and
logistically centralised in terms of the supported services and shared
resources that enable a coordinated functionality in support of SKA science.

Such a persistent and coherent set of infrastructure and services can
only be successfully achieved through an enabling organisational
structure, that will be responsible for the coordination of the
distributed development and operational effort. In order to meet these
requirements, the individual SRCs cannot be independent units, but
will need to work together as a {\it SRC network} to jointly deliver
the functionality needed by the SKA user community. Each SRC is likely
to differ greatly in scale, timeline and financial resources
available. The interests, stakeholders and priorities, of each SRC may
also be different but each needs to contribute to a converged,
coherent, and logistically centralised international SRC network that
meets the needs of an operational observatory, while being
simultaneously responsive to the needs of key projects and the teams
that will run them. So although the development of the SKA network
will be a completely decentralised process in many respects, the
capability and functionality of the network needs to defined and
coordinated by a common entity.

Another aspect became clear in the early investigations by the
SRCCG. Given the lack of experience in the astronomy community in
building a data centre network on this scale, capable of collectively
ingesting and processing 600-700 PB of data every year, it was felt
that a step-by-step approach would have many benefits. In the first
step, a smaller version of the SRC called a "proto-SRC" could be built
in each of the regions which are hoping to host a SRC in the future. The
proto-SRCs are expected to provide most of the functionality of a full
SRC, but on a much smaller scale of compute and storage
capacity. These proto-SRCs would be very useful to:

\begin{itemize}
    \item study and overcome the multifarious technical challenges associated with development of large data centres
    \item develop the human expertise required to set up and run an SRC. Future science users of the SKA could also be trained in the proto-SRCs, on how the new paradigm of data analysis is to be used in the SKA era. 
    \item process data from the various SKA pathfinders and precursors at scale. These proto-SRCs are also expected to develop and test new algorithm and pipelines for the exascale data analysis needed for SKA data. 
    \item offer compute, storage and visualisation resources to users who wish to carry out numerical simulations of relevance to SKA science.
\end{itemize}

Between 2016-2018, the SRCCG, carried out work on defining some of the
basic requirements \citep{skareq} and challenges for a network of SRCs
and began to work with SRC efforts in individual SKA member countries as they
sought to create proto-SRC projects. By the end of 2018, proto-SRC
design and development projects were in advanced stages of planning
and initiation across 13 SKA member states.  This changed ground
situation necessitated parallel changes in the SRCCG, to a new body that
could take this work forward from the planning stage to the
proto-implementation stage. The new body would also be charged with
coming up with a detailed implementation plan for the SRC
network. Accordingly, in November 2018, the SKA Board approved the
formation of the SRC Steering Committee (SRCSC). The mission of the
SRCSC was defined as: "Guide the definition and creation of a
long-term operational partnership between the SKA Observatory and an
ensemble of independently-resourced SKA Regional Centres." As a first
step in implementing this broad mandate, the SRCSC produced a white
paper \footnote{https://aussrc.org/wp-content/uploads/2021/05/SRC-White-Paper-v1.0-Final.pdf}
to jointly define the function and form of an operational SKAO/SRC
collaboration. Thereafter, the SRCSC reorganised itself into six
working groups who worked to define requirements on the software,
hardware, network and operations for the future SRCs. These
requirements were based on extensive consulations with the astronomer
user community. Each requirement was carefully analysed and reviewed,
within each working group and across working groups. It is commendable
that all of this work was carried out completely online through the
Covid-19 pandemic. The next step is to develop a small selection of
prototypes in different areas of SRC development. This work will
require considerable reorganisation of the working group structure
which has commenced in early 2022. The eventual goal of the SRCSC is
to develop a detailed SRC network implementation plan by the middle of
2023. The plan will outline the high level structure of the SRC
network, the main technological challenges and how they are to be
overcome, and propose a governance structure for the efficient working
of the network and a financial plan with some details on which
resources would be available. More than a hundred scientists and
engineers from across all SKA member countries are involved in this
effort.

The next section summarises some aspects of the shared understanding within the international SKA SRC community on the contours of the future SRC network.

\section{The SKA Regional Centre network}

It is likely that there will be about half a dozen SRCs distributed
across the world. One possible layout of the SRCs in shown in Figure
1. Each SRC will need to be connected to every other SRC and to the
two telescope sites located on two different continents.  Undersea
high-speed optical fibre networks will play a crucial role in
providing this capability.

The further challenge is that the SRCs will not be a part of the SKAO,
nor be directly funded by it, but are nevertheless essential to
generate scientific results from the observatory. The SRCs will be
coordinated with assistance from SKAO and will be accredited with the
SKAO. The principle functions of the SRCs will be:

\begin{itemize}
    \item Take data products generated by the science data handling and processing within the SKAO and turn them into science ready data products for astronomers to use
    \item Support (with hardware, software and user support) regional astronomers with their data processing
    \item Act as a centre for domain expertise to enable maximal science with SKAO data and any computational work relevant to SKAO related science.
\end{itemize}

\subsection{What each SRC will do}

Each SRC will provide a nexus for resources - combining scientific
expertise, software expertise and access to computing hardware. In the
software arena, they will provide support for development of
subject-specific pipelines, specific to the key science projects led
by scientists in their region. Each SRC will contribute hardware and
software resources to common efforts in visualisation and creation of
value added data products through stacking, co-adding etc. They will
provide resources to run, access and visualise SKA simulations. For
the SKA data itself, they will provide access to data while
simultaneously ensuring security and adherence to SKA data
policies. They will play a defined role in hosting and distributing
archives. Since the SRCs are likely to be well distributed in
longitude, they will also provide local (time zone) user support,
proposal access, information, training and support for outreach
activities. They will act as a liaison between the SKA Observatory and
the National Research and Education Networks (NRENs), so that
sufficient and affordable network capacity is procured and provisioned
in a timely fashion.

\subsection{Unified data access through the SRC network}

Data from the twin SKAO telescopes will flow via dedicated optical
fibres to their respective science data processing centres in Perth and Cape
Town, and from there onwards to the worldwide network of SRCs
\citep{techinterface}.  Network bandwidth procured and managed by the
National Research and Education Network (NREN) of each SRC host
country will be utilised for transporting data internationally and
within each SRC host country. This interconnected network will provide
a unified science platform to the SKA user community. SKA users will
login to a common science portal. The data flow from telescope host
countries to the SRCs and onwards to the users is visualised in
Fig.~\ref{fig:srcworkflow}. After authentication, they will gain access
to various SKA datasets for which they are authorised. Some data will
be accessible to all, while other data may be restricted to the key
science project group or PI project group to which a particular user
belongs. The user will be permitted to combine different datasets from
the SKAO as well as images, catalogs and spectra from other telescopes
and process them together using a customised workflow designed for a
particular scientific goal. Throughout, the user need not worry about
where her data are sitting or where compute resources are being put to
use to execute the analysis she needs.  Such seamless integration of
storage, compute and network will be imperative to maximise the
scientific return from the SKA.

It is expected that users from all SKA member countries will access
the collective resources of the SRCs via a single sign-on on a common
science portal with appropriate authentication and
authorisation. Users will be able to run queries that require data to
be combined from multiple SRC sites. The movement of data across SRCs
in response to a user's query will remain completely invisible to the
user. Users will also have access to computational resources for
carrying out further data analysis. The extent of this availability
will be dependent on their membership of key science projects and
other collaborations, in accordance with extant SKAO policies.

The next section describes specific plans for a proto-SRC in India.

\section{A proto-SRC for India}

\subsection{Why build an SRC in India?}

India already has a large community of potential SKA users and
attempts are being made to grow these numbers and to train existing
members for using the SKA (See article on Human capacity development
in this volume).  Indian scientists will play a leadership role in
several Key Science Projects (KSPs) which will need significant SRC
resources.  By becoming a part of the worldwide development effort by
contructing our own SRC, we can reduce duplication of effort in both
software and hardware development, and thus costs. Experience gained
while building Exascale infrastructure and systems for the SRC will be
relevant to other areas of data-intensive research across the physical
sciences and elsewhere.  These Exascale problems also demand novel
methodological approaches. For this, the development of Artifical
Intelligence and Machine Learning based methods will be crucial. With
an SRC in India, we will be able to leverage existing expertise in
these areas, from academic research institutions as well as industry.
Feedback will also flow in the opposite direction. Through the SRC,
expertise in data intensive (Exascale) computing will be nurtured in
academic institutions as well as in industry. The Exascale challenges
of the SKA provide a unique and valuable training ground for future
data scientists who will grow Indian capabilities in data
science. Such a symbiotic partnership with multiple stakeholders can
be expected to provide benefits well beyond astronomy, in the long
term.

The location for the Indian SRC is not finalised. However, given the
requirements of good network connectivity, reliable power supply and
availability of trained manpower to build and maintain a large data
centre, the options for housing it become strongly constrained. As a
baseline proposal, an SRC at the NCRA-TIFR campus in Pune with its own
dedicated space in an independent, custom designed building, has been envisaged.

\subsection{Proto-SRC capabilities}

As discussed earlier, the construction of such large data centres is a
technical challenge for all SKA member nations. To overcome this
challenge, each country plans to start small by constructing a smaller
version of the SRC over the next 5 years, known as a proto-SRC. In
India, we propose to take the same approach and construct a proto-SRC
over the next four years.  The proto-SRC will be housed in the
existing data centre at the NCRA-TIFR campus in Pune, India. The
proto-SRC will evaluate emerging hardware and software technologies
and also begin software development in areas of relevance to SKA data
processing and analysis such as algorithm implementation, pipeline
development, and data visualisation software. All of these
developments will be carried out in close cooperation with similar
developments in other SKA member countries, so that software and
learnings can be widely shared for mutual benefit.

We envisage that the Indian proto-SRC will be used in three major modes:
\begin{enumerate}
    \item Data centre capabilities will be used to process and analyse large data sets from the SKA precursor and pathfinder telescopes which have strong Indian participation such as uGMRT, MWA and  MeerKAT. Each of these telescopes will produce datasets of a Petabyte or larger and can benefit greatly from the availability of a large data centre with shared infrastructure.
    \item As the SKA construction proceeds, data will be available to us from the SKA early array releases from about 2024 onwards. These datasets will also be processed in the proto-SRC to gain experience with SKA data processing and also to carry out early science in areas of interest to the Indian community.  
\item  The SKA will explore aspects of the Universe that have never been studied before.  In such a situation, advanced computer simulations that use known physics to make predictions about the properties of the Universe that are only observable with SKA, will be critical to inform the observations and data analysis. Indian astronomy groups have developed such complex simulations over the last few decades. These simulations require large amounts of compute and storage resources. The prototype regional centre will aim to, at least partially, cater to these requirements.  
\end{enumerate}

\subsection{Proto-SRC storage}

We expect to build up a storage capability of $\sim$10 Petabytes with
commensurate compute for data analysis and simulations over a four
year period (2022-26).  The facility will be built up over time and
will provide opportunities to evaluate the performance of emerging
hardware and networking technologies for possible eventual use in the
full SRC. The storage solution will be procured from leading vendors
after carefully comparing performance of prototype systems from
different vendors. We expect that the system will be a combination of
SSD and HDD based storage, with the exact proportion of each being
determined via a tradeoff between cost and performance. Emerging fast
storage options like NVMe will also be evaluated for possible
inclusion in our storage stack.

\subsection{Proto-SRC compute}

The compute requirements will be driven by the combined requirements
of data processing (from SKA early array releases and SKA
pathfinders/precursors) and the simulations. We will use a mix of GPU
and CPU based compute hardware, with a view to measure their
performance in a real-life applications. The optimal ratio of CPU to
GPU capacity in the compute stack will be determined after careful
evaluation of the performance of each on real life radio astronomy
data. We will work directly with hardware vendors and try to borrow
new hardware for testing purposes, so that the most efficient
configuration can be identified and then procured. We will also
consult our colleagues who are building proto-SRCs in other countries
and learn from their experiences.

During 2022-23, we aim to buy server class CPU-based compute hardware
with about 20 TFLOPS capability. Additional add-on cards for GPU
computing will be added to some of these servers. This compute
hardware will be used for about 5 years, before it is replaced with
the compute hardware of the full SRC.

\subsection{Proto-SRC network}

The proto-SRC will need to offer high speed network connectivity to
its users, since most users will be located offsite and distributed
across the country. The National Knowledge Network (NKN) functions as
the NREN for India. The NKN is a multi-gigabit national research and
education network, whose purpose is to provide a unified high speed
network backbone for educational and research institutions in
India. It is managed by the National Informatics Centre. The NKN now
connects over 1600 research institutions, universities and colleges
with its state-of-the-art network across the country. Appropriate
network speed is made available to each NKN connected institution,
depending upon its usage pattern. If usage increases, additional
bandwidth can be added. We expect that since SKA India Consortium
members are part of the NKN, their network bandwidth requirements for
their usage of the proto-SRC will be adequately met through the
NKN. Some institutions may require an upgrade in their bandwidth
connection to the NKN.

\subsection{Proto-SRC sofware development}

Various softwares that will run on the proto-SRC is a critical
component of the ecosystem. An open-source software strategy, which is
already in wide use in astronomy, will be adopted. This will allow for
active and continuous cooperation with such efforts in other
countries. The FAIR Data Principles (Findable, Accessible,
Interoperable, and Reusable) will be adopted in the software we
develop.  Software that could be developed in the proto-SRC could
include (but is not limited to) data analysis pipelines, algorithm
implementation, virtual observatory interfaces, searchable meta-data
archives, middleware, user interfaces, cloud computing tools and data
visualisation software.

We aim to hire about 4 software developers to work full-time on the
SRC software effort. Additional contributions of person power from SKA
member institutions for software development will be actively
solicited and welcomed. The Indian software team will work closely
with their counterparts in other SRC-host nations to collaboratively
develop the numerous software components that will be needed for a
functional proto-SRC. The proto-SRC software will be designed in such
a way that it will scale effectively to the full SRCs, when those are
constructed in the future. A scaled agile framework (SAFe) based
software development methodology will be adopted for the software
development.

\subsection{The proto SRC partnership at the national level}

For the success of the proto-SRC, it is crucial that it be developed
as a partnership between the academic community, the industry partners
and the NKN. The academic research organisations who are part of the
SKA India Consortium will be responsible for design, procurement,
integration testing. They will be responsible for astronomical
pipeline prototyping and user support. The industry partners will be
suppliers of CPU and GPU servers and storage systems. They will be
charged with maintenance of all hardware systems and development of
software tools for data centre management. They will carry out data
pipeline profiling, optimisation and refactoring and work towards
scaling up of the software pipelines developed in the academic
environment to factory scale. They could also supply skilled
developers to develop data visualisation software for SKA
data. Finally, NKN assistance will be needed for maintaining high
speed data links with all proto-SRC users and with other proto SRCs
across the world and with the SKAO telescopes.

\subsection{Collaboration with other proto-SRC efforts}

All proto-SRCs face similar technical challenges. We expect that
there will be extensive collaboration with other efforts across the
world. This collaboration will include joint software development as
well as sharing of knowledge and learnings on new compute and network
hardware. Collaboration can be enabled and coordinated via the SRC
Steering Committee in the short term and by the SRC network entity, as
and when it comes into existence.

Working groups have been already been setup by the SKAO to define the
requirements of the SRC and to develop an architecture for each SRC
and for the SRC network. Efforts are presently underway to develop a
few prototypes that implement some of the new technologies that seem
promising for use in the future SRCs. This is being done by setting up
teams that will work together within the scaled agile framework, in
rapid development cycles in three month long program
increments. Significant Indian contribution to this effort from both
academia and industry will be important; in order for India to benefit
fully from the efforts of the international community.

\section{Future plans - the full SRC}  

In the second half of the decade, we will take up the construction of
the full SRC. The full SRC will require the construction of a data
centre building to house the facility. The full SRC will have storage,
compute and network resources which will be shared with other
SRCs. The resources will be commensurate with the data being produced
by both SKA telescopes \citep{load,cost}.

Our SRC personnel will collaborate closely with other countries
housing an SRC, in the design and development process, to ensure that
the Indian SRC is seamlessly connected to the SRC network.  We will
time the construction of the full SRC so that it is ready for use as
soon as the SKA Phase 1, commences operations. Our SRC will require
high-speed network connectivity to the other SRCs and with the two
telescope sites. We will also need to provide fast network
connectivity to users to enable them to access and process SKA data on
the SRC.

The experiences and learnings from the Indian proto-SRC and from other
proto-SRCs across the world and the design, development and prototype work
currently underway in the SKAO-guided international collaboration will
play a crucial role in defining the contours and specifications of the
future Indian SKA Regional Centre.

\section*{Acknowledgements} 

YW thanks all the members of the SRCSC and the SRCCG for numerous
discussions over the last five years that have helped seed and develop
many of the ideas presented in this paper. YW and DO acknowledge the
support of the Department of Atomic Energy, Government of India, under
project no. 12-R\&D-TFR5.02-0700. SP acknowledges the support of
Department of Science and Technology (code: IF-12/PH-44) and Science
and Engineering Research Board (SR/FTP/PS-118/2011).

\begin{figure}
\includegraphics[angle=0,width=9.0cm,trim={0.0cm 0.0cm 0.0cm 0.0cm},clip]{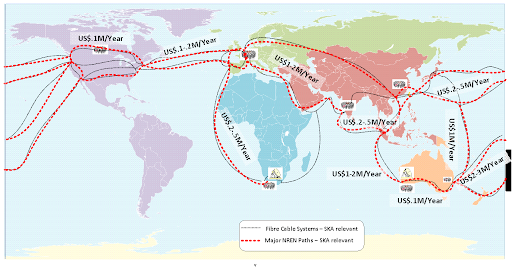}
\caption{Possible locations of SKA regional centres across the world. SKA relevant fibre cable systems are indicated, along with major NREN paths and the cost per year of a 100 Gbit 10-year Indefeasible Right of Use (IRU) contract. Image Credit: SKA Regional Centre Steering Committee}
\label{fig:srclayout}
\end{figure}

\begin{figure}
\includegraphics[angle=0,width=9.0cm,trim={0.0cm 0.0cm 0.0cm 0.0cm},clip]{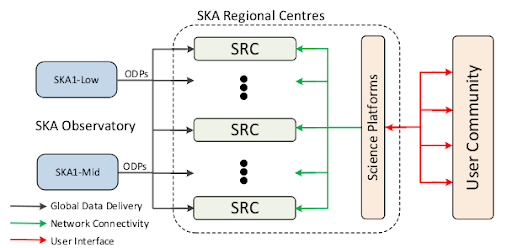}
\caption{Data flow diagram from the telescope sites to the SRCs and onward to the user community via the common science platform. Image Credit: SKA Regional Centre Steering Committee}
\label{fig:srcworkflow}
\end{figure}

\end{document}